\documentclass[aps,pre,twocolumn,showpacs,groupeaddress,floatfix]{revtex4-1}
\usepackage{graphicx,bm,amsmath,amssymb,natbib,url,epsfig}
\usepackage{dcolumn}
\usepackage{hyperref}
\usepackage{amsmath,amssymb}

\newcommand{\ohm}{$\Omega$}

\begin{document}


\title{Control of birhythmicity through conjugate self-feedback: Theory and experiment} 



\author{Debabrata Biswas${}^1$}
\email{debbisrs@gmail.com}
\author{Tanmoy Banerjee${}^{1}$}
\email{tbanerjee@phys.buruniv.ac.in}\thanks{(Author for correspondence)}
\author{J\"urgen Kurths${}^{2,3,4,5}$}%
\email{juergen.kurths@pik-potsdam.de}
\affiliation{%
${}^1$ Chaos and Complex Systems Research Laboratory, Department of Physics, University of Burdwan, Burdwan 713 104, West Bengal, India.\\
${}^{2}$ Potsdam Institute for Climate Impact Research, Telegraphenberg, D-14415 Potsdam, Germany.\\
${}^3$ Institute of Physics, Humboldt University Berlin, D-12489 Berlin, Germany.\\
${}^4$  Institute for Complex Systems and Mathematical Biology, University of Aberdeen, Aberdeen AB24 3FX, UK.\\
${}^5$ Institute of Applied Physics of the Russian Academy of Sciences, 603950 Nizhny Novgorod, Russia.}%


\received{:to be included by reviewer}
\date{\today}

\begin{abstract}
Birhythmicity arises in several physical, biological and chemical systems. Although, many control schemes are proposed for various forms of  multistability, only a few exist for controlling birhythmicity. In this paper we investigate the control of birhythmic oscillation by introducing a self-feedback mechanism that incorporates the variable to be controlled and its canonical conjugate. Using a detailed analytical treatment, bifurcation analysis and experimental demonstrations we establish that the proposed technique is capable of eliminating birhythmicity and generates monorhythmic oscillation. Further, the detailed parameter space study reveals that, apart from monorhythmicity, the system shows transition between birhythmicity and other dynamical forms of bistability. This study may have practical applications in controlling birhythmic behavior of several systems, in particular in biochemical and mechanical processes. 
\end{abstract}


\maketitle 

\section{Introduction}
\label{sec:intro}
Multistabity is a common dynamical feature of many natural systems \cite{gold,goldnat,pf}. Although it appears in diverse forms, a very frequently occurred variant is bistability. There are three main manifestations of bistability: The coexistence of  (i) two stable steady states, (ii) one stable limit cycle and one stable steady state, and (iii) two stable limit cycles. The third form of bistability, i.e., the coexistence of two stable limit cycles of different amplitude and frequency, generally separated by an unstable limit cycle, is called birhythmicity and oscillators showing this behavior are called birhythmic oscillators. Apart from two coexisting periodic limit cycles, birhythmicity may appear in a much more complex form, e.g., coexistence of two chaotic attractors \cite{pis_prl,chua_co}. Birhythmic oscillators are very common, particularly, in physics (e.g., energy harvesting system, see Ref.~\cite{eh} and references therein) biology (e.g. glycolytic oscillator and enzymatic reactions \cite{gold,goldnat,karroy04}) and chemistry \cite{epst}. Most of the biochemical oscillations that govern the organization of cell cycle, brain dynamics or chemical oscillations are birhythmic; examples include, birhythmicity in the p53-Mdm2 network, which is the key protein module that controls proliferation of abnormal cells in mammals \cite{pone,natp53}, intracellular Ca$2+$ oscillations \cite{goldnat}, oscillatory generation of cyclic AMP (cAMP) during the aggregation of the slime mold {\it Dictyostelium discoideum} \cite{gold_cAMP,*nat_cAMP}  and circadian oscillations of the PER and TIM proteins in {\it Drosophila} \cite{gold_cir}. 

In physical and engineering systems birhythmicity plays a negative role in limiting the efficiency of a certain application. Take the practical example of an energy harvesting system that converts wind-induced vibrational energy into electrical energy. This type of energy harvesting systems show birhythmicity \cite{eh}, but for an efficient harvesting it is desirable that the system always resides on the large amplitude limit cycle because that produces a significant mechanical deformation, which, in turn results in larger amount of harvested electric power. Further, the presence of birhythmicity makes a system vulnerable to noise: depending upon the noise intensity the system may end up in any of the two limit cycles, which results in an unpredictable system dynamics \cite{pf,noise}. Therefore, monorhythmicity is of practical importance in most of the physical systems. On the other hand, in networks of neuronal oscillators the occurrence of birhythmicity is often desirable to generate and maintain different modes of oscillations that organize various biochemical processes in response to variations in their environment \cite{epst}. Therefore, identifying an efficient control technique is of importance that can tame birhythmicity to yield monorhythmic oscillation or can retain its character intact where ever needed.

Although several mechanisms are proposed for controlling bistability consisting of oscillation and steady state \cite{pg,*pis01,*pg2} (for an elaborate recent review on the control of multistability see \cite{pf} and references therein), only a few exist to control birhythmicity. \citet{Ghosh11} reported an effective control mechanism of birhythmic behavior in a modified van der Pol system by using a variant of Pyragas technique of time delay control \cite{Pyragas95} and they showed that depending upon the time-delay one can induce monorhythmic oscillation out of birhythmicity. But, due to the presence of time delay the system becomes infinite dimensional and thus a detailed bifurcation analysis for a wide parameter space is difficult and was not reported there. Further, the authors of \cite{Ghosh11} established that their technique can {\it suppress} the effective birhythmic zone but can not eliminate it completely for all possible sets of nonlinear damping parameter values. In this context, another interesting control technique has been reported recently by \citet{pis_rsca}, where the authors demonstrated that multistable systems with coexisting either periodic or chaotic attractors can be converted into a monostable one by applying an external harmonic modulation and a positive feedback to a proper {\it accessible} system parameter.

In the present paper we propose an effective and much more general control technique, that we call the conjugate self-feedback control, which is able to eliminate birhythmicity and induce monorhythmic behavior. We consider a modified van der Pol equation that has been proposed to model enzyme reactions in some biosystems \cite{kaiser1,kaiser2} and also has been studied earlier as a prototypical model that exhibits birhthmicity \cite{kaiser1,kaiser2,Ghosh11}. With a detailed bifurcation study we establish the effectiveness of the proposed control technique in taming birhthmicity and inducing monorhythmicity. Depending upon the value of the self-feedback strength it also offers freedom to select one of the desired dynamics. We also demonstrate our results experimentally with an electronic circuit and verify that our results are robust enough in a real-world setup where the presence of parameter fluctuation and noise are inevitable.

\section{The van der Pol oscillator with birhythimicity}
\label{sec:uncoun}

First we describe the model used in the following. Consider a birhythmic van der Pol oscillator given by \cite{kaiser1,kaiser2}
\begin{equation}
\label{vdpeq}
\ddot{x}-\mu(1-x^2+\alpha x^4-\beta x^6)\dot{x}+x=0.
\end{equation}
Here, $\mu>0$, $\alpha>0$ and $\beta>0$ are parameters that determine the nonlinear damping. 

\begin{figure}
\centering
\includegraphics[width=0.40\textwidth]{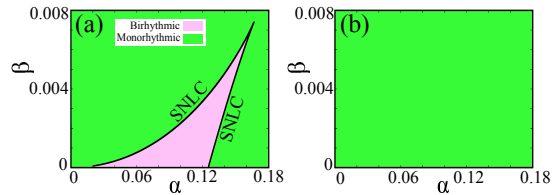}
\caption{The two parameter bifurcation diagram in the $\alpha-\beta$ space with $\mu=0.1$. Bifurcation diagram for (a) Eq.~\eqref{vdpeq} (i.e., without control), SNLC: Saddle-node bifurcation of limit cycle; (b) With control (for $d=-0.1$ of Eq.\ref{vdpeqcoup}): the control eliminates birhthmicity and only one limit cycle (LC) exists. }
\label{ab}
\end{figure}

In Ref.~\cite{Kadji07}, Kadji {\em et al.} considered, $x(t)=A\cos\omega t$, and by using the harmonic decomposition method they arrived at the following amplitude equation: 

\begin{equation}
\label{ampeqa}
\frac{5\beta}{64} A^6-\frac{\alpha}{8} A^4+\frac{1}{4} A^2-1=0.
\end{equation}
Equation \eqref{ampeqa} is the generic form of the codimension-two saddle-node (SN) bifurcation. Note that Eq.~\eqref{ampeqa} is independent of the parameter $\mu$. The two parameter bifurcation diagram in the $\alpha-\beta$ parameter space is shown in Fig.~\ref{ab} (a) that exhibits a cusp type of codimension-two bifurcation. The exchange of rhythmicity is through the saddle-node bifurcation of the limit cycle (SNLC) (shown by the solid black line in the figure). Fig.~\ref{ab}(b) shows the controlled case with $d=-0.1$ ($d$ is the control parameter to be discussed later), where birhythmicity is completely removed and only monorhythmicity exists. 

\section{Control of birhythmicity through conjugate self-feedback: Theory}
\label{sec:coup}
Next we introduce a conjugate self-feedback term $d(\dot{x}-x)$ in Eq.~\eqref{vdpeq}
\begin{equation}
\label{vdpeqcoup}
\ddot{x}-\mu(1-x^2+\alpha x^4-\beta x^6)\dot{x}+x+d(\dot{x}-x)=0,
\end{equation}
which contains the variable of our interest, $x$, and its canonical conjugate \cite{clbook}, $\dot x$; here $d$ controls the strength of the self-feedback. Further, a close inspection reveals that the self-feedback mechanism effectively controls the damping of the system through the $\dot x$ variable and the effective frequency through the $x$ variable. However, understanding of their collective effect on the dynamics needs a detailed analysis that we will address next.   

To unravel the underlying dynamics of the controlled system we use the harmonic decomposition method. Let us assume the approximate solution of \eqref{vdpeqcoup} be given by
\begin{equation}
x(t)=A\cos\omega t,
\end{equation}
with $A$ being the amplitude and $\omega$ the frequency of the oscillator with feedback. Substituting this in \eqref{vdpeqcoup} yields the following expression
\begin{equation}
\label{coupamp}
\begin{split}
\big(-&\omega^2-d+1\big)A\cos \omega t=\\
&-\mu\omega\bigg(1-\frac{1}{4}A^2+\frac{\alpha}{8}A^4-\frac{5\beta}{64}A^6\bigg)A\sin\omega t\\
&+d\omega A\sin\omega t\\
&+\mu\omega\bigg(\frac{1}{4}A^2-\frac{3\alpha}{16}A^4+\frac{9\beta}{64}A^6\bigg)A\sin 3\omega t\\
&-\mu\omega\bigg(\frac{\alpha}{16}A^4-\frac{5\beta}{64}A^6\bigg)A\sin 5\omega t\\
&+A^7\beta\mu\omega\sin 7\omega t.
\end{split}
\end{equation}
But according to Ref.~\cite{Jordan99}, we can ignore the higher harmonics regarding them as forcing term, which diminish with increasing harmonics. Thus, Eq.~\eqref{coupamp} can be reduced to
\begin{equation}
\label{coupamp1}
\begin{split}
\big(-\omega^2&-d+1\big)A\cos \omega t=\\
&-\mu\omega\bigg(1-\frac{1}{4}A^2+\frac{\alpha}{8}A^4-\frac{5\beta}{64}A^6\bigg)A\sin\omega t\\
&+d\omega A\sin\omega t+\text{higher harmonic terms}.
\end{split}
\end{equation}
The equation \eqref{coupamp1} suggests the following frequency and amplitude equations, respectively,
\begin{equation}
\label{freq}
\omega^2+d-1=0,
\end{equation}
and
\begin{equation}
\label{amp}
\mu\bigg(1-\frac{1}{4}A^2+\frac{\alpha}{8}A^4-\frac{5\beta}{64}A^6\bigg)-d=0.
\end{equation}
It is interesting to note that, Eq.~\eqref{amp} is equivalent to Eq.~\eqref{ampeqa} when $d=0$, i.e., in the absence of any feedback. Also, it may be noted that the amplitude of the system depends on $\mu$ when $d\neq 0$, contrary to Eq.~\eqref{ampeqa}. The frequency in the harmonic limit corresponds to $\omega=1$. Further, Eq.~\eqref{freq} imposes an upper limit on the strength of the feedback, namely $d\le1$ otherwise the frequency becomes imaginary, which is non-physical. The three roots (actually six roots, with $\pm A_i$, $(i=1,2)$.) correspond to the amplitudes of three limit cycles (two stable, one unstable).
\begin{figure}
\centering
\includegraphics[width=0.40\textwidth]{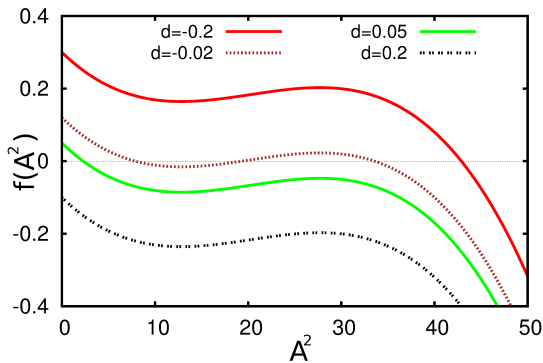}
\caption{Plot of $f(A^2)-A^2$ for the parameter set $\mu=0.1$, $\alpha=0.114$, $\beta=0.003$ for different values of the coupling parameter $d$. The solid dark (red) curve for $d=-0.2$ represents single limit cycle with large amplitude and the solid gray (green) curve for $d=0.05$ represents that with small amplitude. In between the curve for $d=-0.02$ is for birhythmic oscillation. The lower curve for $d=0.2$ represents stable steady state.}
\label{ava}
\end{figure}
We can get a hint of the amplitude of the limit cycles and test the stability using the energy balance method as suggested in Ref.~\cite{Ghosh11}. From Eq.~\eqref{vdpeqcoup} one can infer that, for $\mu=0$ and $d=0$, the harmonic solution may be given by \cite{Yamapi07}
\begin{equation}
\label{harsol}
x(t)=A\cos(t+\phi),
\end{equation}
where, $\phi$ is the initial phase, preferably $\phi=0$ for convenience. The phase plane of this solution is a circle with period $T= 2\pi$. In the presence of self-feedback we can approximate
\begin{equation}
\label{harsold}
x(t)\backsimeq A\cos t.
\end{equation}
Now, the change in energy $\Delta E$ in one period $0\leq t \leq T$, where $T=2\pi$, may be found out if one considers the term $\mu(1-x^2+\alpha x^4-\beta x^6)-d(\dot{x}-x)$ as the external forcing term by the following way
\begin{eqnarray}
\Delta E&=&E(T)-E(0),\nonumber\\
&=&\int_0^T\big[\mu(1-x^2+\alpha x^4-\beta x^6)-d(\dot{x}-x)\big]\dot{x}dt.~~~~~\label{int}
\end{eqnarray}
For a periodic solution (limit cycle), the change in energy must be zero, i.e., $\Delta E=0$. Hence the above integration along with the condition of Eq.~\eqref{harsold} yields
\begin{eqnarray}
f(A^2)=\mu\bigg(1-\frac{1}{4}A^2+\frac{\alpha}{8}A^4-\frac{5\beta}{64}A^6\bigg)-d=0.\label{fa}
\end{eqnarray}
Again, we see that, Eq.~\eqref{fa} is identical to Eq.~\eqref{amp}. In the absence of the coupling Eq.~\eqref{fa} reduces to Eq.~\eqref{ampeqa}. The saddle-node bifurcation may be controlled by changing the value of $d$. Eq.~\eqref{fa} may be solved to have a number of positive roots, which determines the number of limit cycle (LC). One can determine the stability of the limit cycle by the slope of the curve of Eq.~\eqref{fa} at the zero crossing points. The negative slope determines the stable limit cycle. Thus, we can write
\begin{equation}
\label{dda}
\frac{d\Delta E(A)}{dA}\bigg\lvert_\text{Limit cycle}<0,
\end{equation}
as the condition for a stable limit cycle.

Now let us discuss how to determine the presence of limit cycles and their stability out of the above analytical results. The amplitude equation Eq.~\eqref{fa} may be solved by graphical method. The solutions are those for which the function $f(A^2)$ crosses the horizontal zero line. We consider the parameter set $\mu=0.1$, $\alpha=0.114$ and $\beta=0.003$ for which \eqref{vdpeqcoup} exhibits birhythmicity in the absence of self-feedback; next, we vary the coupling strength $d$ to get different solutions. The number of limit cycles is determined by the number of solutions of the amplitude equation. The number provides the information of the steady state solution (i.e., no solution), existence of a single limit cycle (monorhythmicity) or three limit cycles (birhythmicity, one of the LCs is unstable). From Fig.~\ref{ava} we find that for $d=0.2$ there is no zero crossing of the curve, i.e., there is no LC and the system is in a steady state. As we decrease $d$, the $f(A^2)$ curve crosses the horizontal zero line from below and gives rise to a stable LC. This is shown for $d=0.05$ with solid gray (green) line, here the system has only one stable LC of small amplitude. Further decrease in $d$ brings it to the birhythmic regime where the $f(A^2)$ curve crosses the horizontal zero line at three different values of $A^2$ indicating three LCs (shown for $d=-0.02$). The stability of three LCs are determined by Eq.~\eqref{dda}, which suggests that the middle zero point of the curve in Fig.~\ref{ava} represents the unstable LC.  Further increase in $d$ brings the system to a monorhythmic region with the large LC. The case of large single LC for $d=-0.1$ is shown in the upper solid dark (red) line.

The original birhythmic van der Pol oscillator given by \eqref{vdpeq} exhibits only global SNLC type of bifurcation. However, due to the presence of the feedback term in the controlled case (i.e., Eq.\ref{vdpeqcoup}), Eq.~\eqref{ampeqa} is modified to Eq.~\eqref{amp}, and thus the system additionally exhibits local bifurcation, namely Hopf bifurcation. We derive the value of $d$ for which Hopf bifurcation occurs from the eigenvalues of the jacobian of Eq.~\eqref{vdpeqcoup} around the steady state $(x,\dot{x})=(0,0)$. The eigenvalues are given by
\begin{equation}\label{eigen}
\lambda_{1,2}=\frac{1}{2}\bigg((\mu-d)\pm \sqrt{(d-\mu)^2-4(1-d)}\bigg).
\end{equation}
Equation \eqref{eigen} gives the condition of Hopf bifurcation as 
\begin{equation}\label{hopf}
d_{HB}=\mu,
\end{equation}
where $d_{HB}$ is the value of $d$ for which Hopf bifurcation occurs. 

\begin{figure}
\centering
\includegraphics[width=0.45\textwidth]{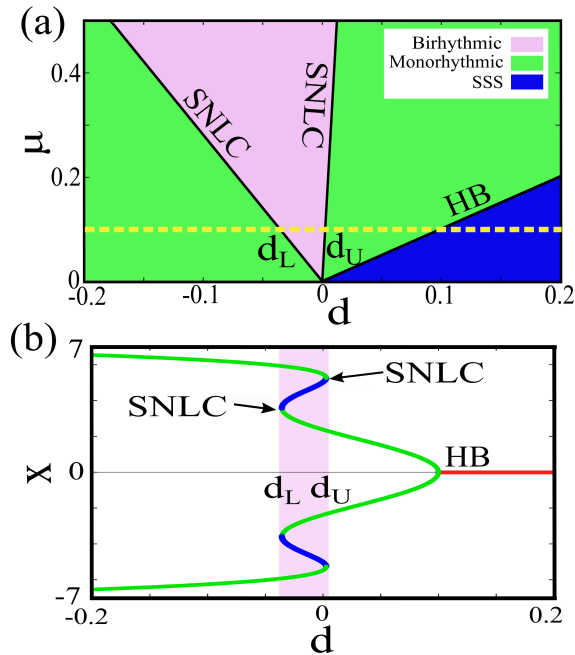}
\caption{(a) Two parameter bifurcation diagram in the $d-\mu$ space for $\alpha=0.114$, $\beta=0.003$, (b) bifurcation diagram with $d$ for $\mu=0.1$ (the horizontal broken line in Fig.~\ref{dmm}(a)). SSS: Stable steady state. ($d_U-d_L$) is the width of birhythmic zone.}
\label{dmm}
\end{figure}
\begin{figure}
\centering
\includegraphics[width=0.41\textwidth]{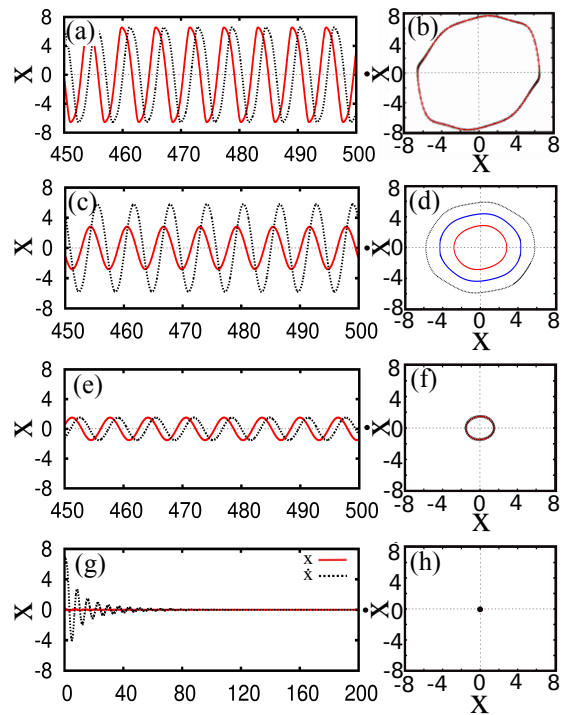}
\caption{Time series and phase plane plots: (a,b) $d=-0.2$: large amplitude single LC. (c,d) $d=-0.02$: Birhythmic oscillations, the blue trajectory in (d) shows unstable LC. (e,f) $d=0.05$: small amplitude single LC. (g,h) $d=0.2$: stable steady state. The solid (red) line is for initial conditions $x_0=0.1$, $\dot{x}_0=0$; the dotted (black) line with initial condition $x_0=7$, $\dot{x}_0=0$.  Other parameters are: $\mu=0.1$, $\alpha=0.114$, $\beta=0.003$.}
\label{vd}
\end{figure}
\section{Numerical Bifurcation analysis}
\label{sec:bifan}
In this section we investigate the possible bifurcation scenarios of the system using the continuation package XPPAUT. We explore the nature of the bifurcation with the variation of the feedback parameter $d$ for different system parameters (e.g., $\mu$, $\alpha$ and $\beta$).  
\subsection{Dynamics in $d-\mu$ space}
The bifurcation structure in the $d-\mu$ space is computed and shown in Fig.~\ref{dmm}(a). The value of $\alpha=0.114$ and $\beta=0.003$ are kept in the birhythmic zone of the uncontrolled system (cf. Fig.~\ref{ab}). We find that the two-parameter space is divided by global bifurcations, namely saddle node bifurcation of limit cycle (SNLC) and a local bifurcation, namely the supercritical Hopf bifurcation (HB). In between two SNLC curves birhythmic behavior exists [purple (gray) zone]: In this zone three LCs exist, of which two are stable (one with smaller amplitude and the other with larger amplitude) and an unstable LC. The transition from birhythmic to monorhythmic dynamics [indicated by green (light gray) zone] is governed by these SNLC curves. Whereas the HB curve governs the transition between single stable limit cycle and stable steady state (SSS) [blue (dark) zone]; note that the occurrence of the Hopf bifurcation agrees with our analytically predicted value of $d$ in \eqref{hopf}.

For a clearer understanding of the bifurcation scenario we take an exemplary value $\mu=0.1$ and vary the feedback term $d$ [along the broken (yellow) horizontal line in Fig.~\ref{dmm}(a)]. The one parameter bifurcation diagram corresponding to this variation is shown in Fig.~\ref{dmm}(b). In the absence of the self-feedback, i.e., for $d=0$, the system is in a birhythmic zone for any $\mu>0$ (in the present parametric set up). If we increase $d$, for $d>d_U$, the system enters into the monorythmic zone via SNLC bifurcation. Here we observe that the sole limit cycle in the system is the small amplitude LC. This small LC looses its stability through an inverse Hopf bifurcation and gives birth to a stable steady state. In the negative side of $d$, for $d<-d_L$, we again have a monorhythmic region but with a large amplitude limit cycle. Therefore, with a proper choice of the self-feedback strength $d$ one can induce monorhythmic oscillation of smaller ($d>d_U$) or larger ($d<-d_L$) amplitude. Interestingly, a hysteresis appears around $d=0$ having a width of $\Delta d=(d_U-d_L)$ [light gray (purple) of Fig.~\ref{dmm}(b)]. In this range of $d$ the system may end up showing LC of large or small amplitude depending upon initial conditions. Also, the two LCs are separated by an unstable LC [shown in dark (blue) line]. It is worth noting that the width of the hysteresis zone increases with increasing $\mu$.

Typical time series with the variation of $d$ are shown in Fig.~\ref{vd} ($\mu=0.1$, $\alpha=0.114$, $\beta=0.003$). To detect the presence or absence of birhythmicity, we consider a large number of initial conditions of $(x,\dot x)$. However, here we present the results for two different initial conditions only: one around the origin (targeting the small amplitude LC) and the other far from the origin (targeting the large amplitude LC). The  red line (solid) indicates the oscillation corresponding to the initial condition $I_1\equiv(x_0,\dot x_0)=(0.1, 0)$ and the black line (dotted) indicates the oscillation for the initial condition $I_2\equiv(x_0,\dot x_0)=(7, 0)$. We start from a negative $d$ with $d<-d_L$. Fig.~\ref{vd}(a) (time series) and  \ref{vd}(b) (phase plane plot) show the scenario for $d=-0.2$.  Both initial conditions result in the large amplitude LC indicating monorhythmicity. Next, we choose $-d_L<d<d_U$, i.e., the birhythmic region. Figure~\ref{vd}(c) and \ref{vd}(d) show this scenario for $d=-0.02$. The blue trajectory in Fig.~\ref{vd}(d) indicates the unstable LC that separates the basin of attraction of two LCs, i.e., the small LC resulted from $I_1$ and the large LC resulted from $I_2$. Figure~\ref{vd}(e) and \ref{vd}(f) show monorhtyhmic oscillation for $d=0.05$ (i.e., $d>d_U$). Here all the initial conditions go to the smaller amplitude LC. Finally, further increase in $d$ results in the stable steady state [Fig.~\ref{vd}(g) and \ref{vd}(h) for $d=0.2$]. Therefore, with the variation of $d$ we can effectively control the birhythmic nature of the system and can induce monorhythmic oscillation of preferred amplitude.

\begin{figure}
\includegraphics[width=0.43\textwidth]{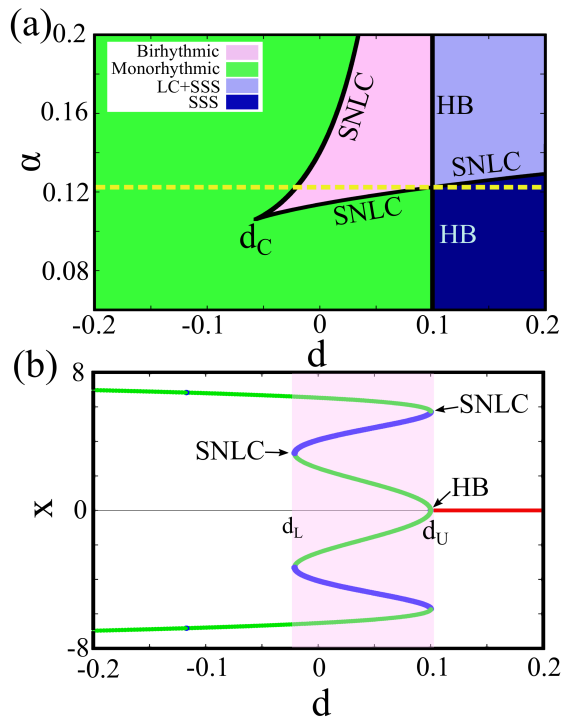}
\caption{(a) Two parameter bifurcation diagram in $d-\alpha$ space for $\mu=0.1$, $\beta=0.003$. The yellow broken line indicates $\alpha_c$ where SNLC and HB curves intersect. $d_C$ is the cusp point. SSS: Stable steady state, LC$+$SSS: bistable zone with one stable steady state and one stable limit cycle. (b) Bifurcation diagram obtained by sweeping $d$ along the  yellow broken line of Fig.~\ref{daa}(a). }
\label{daa}
\end{figure}

\begin{figure}
\includegraphics[width=0.41\textwidth]{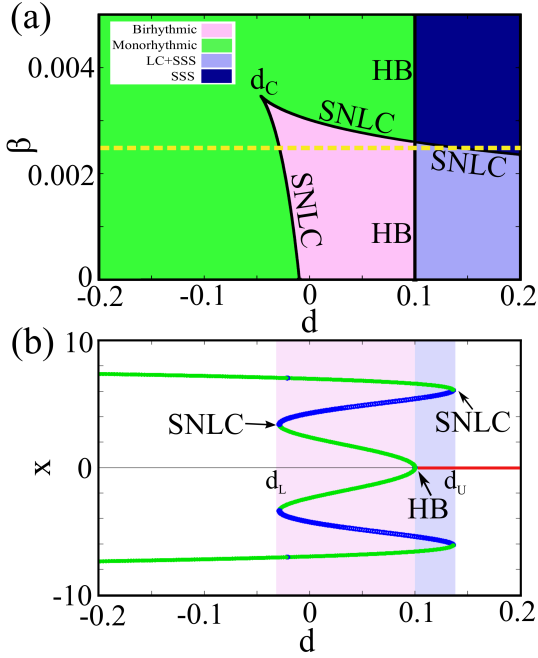}
\caption{(a) Two parameter bifurcation diagram in $d-\beta$ space for $\mu=0.1$, $\alpha=0.114$. $d_C$ is the cusp point. (b) Bifurcation diagram with $d$ for $\beta=0.0025$ [along the yellow broken line of Fig.~\ref{dbb}(a)].}
\label{dbb}
\end{figure}
\subsection{Effect of nonlinear damping parameters}
Next, we investigate the effectiveness of the control over the whole nonlinear damping parameter space. Significantly, we find that one can indeed induce monorhythmicity for any set of ($\alpha$, $\beta$) by choosing a proper value of $d$. To systematically understand the scenario, we study the dynamics in the $d-\alpha$ and $d-\beta$ space, separately. Figure~\ref{daa} shows the two-parameter bifurcation in the $d-\alpha$ space for $\beta=0.003$ and Figure~\ref{dbb}(a) shows the same in the $d-\beta$ space for $\alpha=0.114$ (in both the cases we take $\mu=0.1$). From these two bifurcation diagrams it is seen that for $d<-d_C$ the system has only a single LC for any choice of ($\alpha$, $\beta$) ($d_C$ is the cusp bifurcation point). 

The HB curve and the SNLC curve intersect at $\alpha=\alpha_c$ (say) in Fig.\ref{daa} (a) and at $\beta=\beta_c$ (say) in Fig.\ref{dbb}(a). Figure~\ref{daa}(b) shows the bifurcation scenario with the variation of $d$ along the horizontal broken yellow line of Fig.~\ref{daa}(a) (i.e., for $\alpha=\alpha_c=0.122$). An interesting transition occurs for $\alpha>\alpha_c$ ($\beta<\beta_c$): If $d$ is increased from below, the system generates a transition from birhythmicity to another type of bistability, namely the {\it coexistence of stable LC and stable steady state}. the genesis of this transition is also quite interesting. Normally, in a hysteric transition, the transition from stable steady state to stable LC occurs through a {\it subcritical} Hopf bifurcation and the reverse transition occurs through a SNLC \cite{stbook}, but here two SNLC and one supercritical Hopf bifurcation govern the hysteric transition. This is shown in Fig.~\ref{dbb}(b) for $(\alpha,~\beta)=(0.114,~0.0025)$ by sweeping $d$ along the yellow broken line of Fig.~\ref{dbb}(a). Also note that the Hopf bifurcation occurs at $d_{HB}=\mu=0.1$ and independent of $\alpha$ and $\beta$ as predicted in Eq.~\ref{hopf}.   

Finally, we summarize our results in the $\alpha-\beta$ parameter space. For the uncontrolled system, i.e., $d=0$, birhyhmicity occurs in a broad zone of ($\alpha-\beta$) values as shown in Fig.~\ref{ab}(a). But, for $d<d_C$ the birhythmic zone is completely eliminated and the only possible dynamics is essentially monorhythmic [Fig.~\ref{ab}(a) for $d=-0.1$]. Therefore, our study reveals that a proper choice of the control parameter $d$ can effectively eliminate birhythmicity to establish monorhythmic oscillation and at the same time its variation may give rise to transitions between several interesting dynamical states; by controlling $d$ one can achieve any of these states in a deterministic way.  

\section{Experiment}
Experimental observation of birhythmicity is subtle due to the presence of inherent noise and parameter fluctuation in a real system and also owing to the fact that, in experiments one can record only one oscillation at a time \cite{gold}. The first experimental observation of birhythmicity was made by Decroly and Goldbeter \cite{decgold} in a chemical system, namely the parallel-coupled bromate-chlorite-iodide system. In their experiment the time scale was of the order of few minutes. In biological experimental setups the time scale is usually of the order of few hours, e.g., birhythmic oscillation in the p53 system has two time scales of six and ten hours \cite{p53_expt}. In this context, the experimental observation of birhythmic oscillation in electronic circuit possesses two distinct advantages: first, the time scale is much reduced, of the order of mili second and the second one is the controllability of electronic circuits. 

\begin{figure}
\centering
\includegraphics[width=0.40\textwidth]{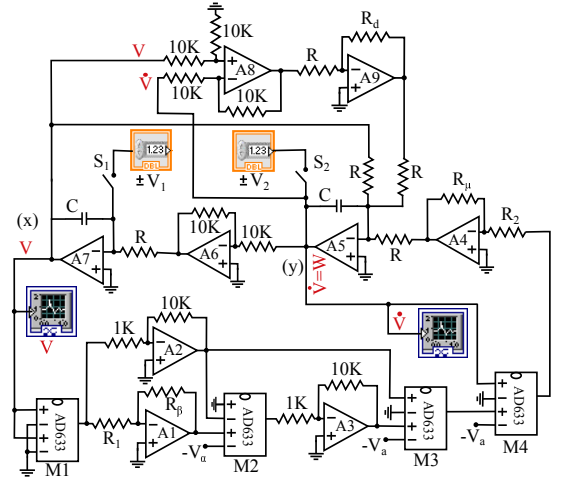}
\caption{The experimental circuit compatible to be controlled and acquired by daq. For description and parameter values see text.}
\label{ckt_init}
\end{figure}

To demonstrate birhythmicity and verify the robustness of our proposed control scheme, we realize the system given by Eq.~\eqref{vdpeqcoup} in the electronic circuit. The detailed circuit diagram is shown in Fig.~\ref{ckt_init}. Here M1-M4 are analog multiplier ICs (AD633JN) and A1-A9 are opamps (TL074). The resulting circuit equation takes the following form
\begin{subequations}\label{syseqckt}
\begin{align}
RC\frac{dV}{dt}&=W,\label{syseqckta}\\
RC\frac{dW}{dt}&=\frac{R_\mu}{100R_2}\bigg[V_a-V^2\bigg(V_a-V^2\bigg(V_\alpha-\frac{R_\beta}{R_1}V^2\bigg)\bigg)\bigg]W\nonumber\\
&~~~-V-\frac{R_d}{R}(W-V).\label{syseqcktb}
\end{align}
\end{subequations}

\label{sec:expt}
The above equation becomes dimensionless for the following substitutions: $t=\frac{t}{RC}$, $x=V$, $y=W$, $\frac{R_\mu}{R_1}=\mu$, $\frac{R_d}{R}=d$, $V_a=1$ V, $V_\alpha=\alpha$ V, and $\frac{R_\beta}{R_1}=\beta$; with these Eq.~\eqref{syseqckt} 
is reduced to Eq.~\eqref{vdpeqcoup}. 

\begin{figure}[t!]
\centering
\includegraphics[width=0.47\textwidth]{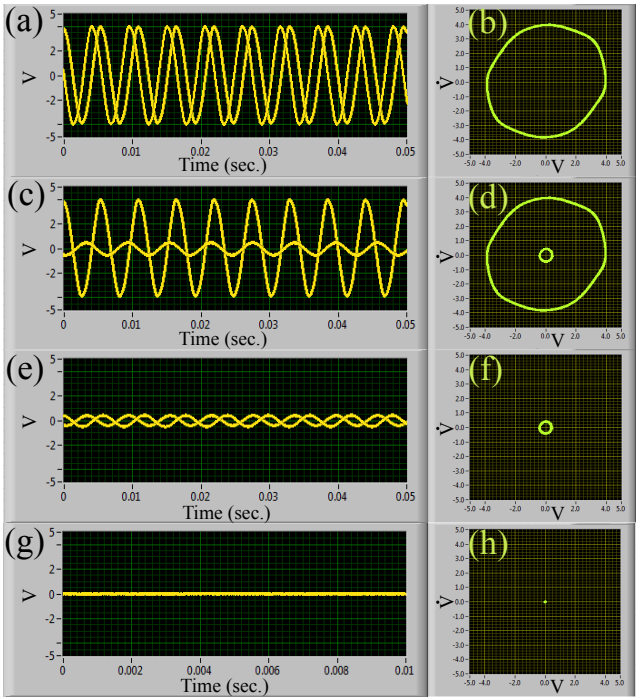}
\caption{The experimental time series and phase plane diagrams obtained by daq. (a,b) $R_d\approx 390$ {\ohm}: large amplitude single LC. (c,d) $R_d\approx 0$ {\ohm}: Birhythmic oscillations. (e,f) $R_d\approx 57.7$ {\ohm}: small amplitude single LC. (g,h) $R_d\approx 895$ {\ohm}: stable steady state. The large amplitude LC is for initial conditions $V_1=2.1$ volt, $V_2=0$ volt and the small amplitude LC is for initial condition $V_1=0.1$ volt, $V_2=0$ volt.}
\label{expt}
\end{figure}

We consider the following values of the used circuit components: $R_\beta\approx 1$ k\ohm, $R_\mu\approx 259.6$ \ohm, $V_\alpha\approx -1.119$ V and $V_a\approx 321.4$ mV throughout the experiment. The initial conditions are controlled through the Data Acquisition System (daq) in Labview environment \cite{labview} through a computer. To have a selected initial conditions, the capacitors ($C$) in the integrators (A5 and A7) are charged with external voltages ($\pm V_1$ and $\pm V_2$). These voltages are controlled by the daq. The voltages are connected to relays (S1 and S2) to be ON for a particular time period. The ON time of the relays are controlled by a microcontroller (Arduino Uno \cite{au}), which is programmed to keep the relays ON for a time interval of $5$ seconds. During this time the capacitors $C$ of the integrators get charged to the desired input voltages ($\pm V_1$ and $\pm V_2$) which are taken and controlled from the computer through the daq. Then the relays are made OFF and the circuit operates in its normal action.

The experimental time series and phase plane plots are shown in Fig.~\ref{expt}. To observe the large amplitude single LC shown in Figure~\ref{expt}(a,b) we add an inverter in the output terminal of A9 of Fig.~\ref{ckt_init} (not shown in the figure) and take $R_d\approx 390$ {\ohm}. Fig.~\ref{expt}(c) and (d) show the scenario of birhythmicity for $R_d\approx 0$ \ohm. The presence of oscillations of two different amplitudes and frequencies confirms the occurrence of birhythmicity in the circuit. The increasing $R_d$ brings the system to a monorhythmic one. The situation for $R_d\approx 57.7$ \ohm~ is shown in Fig.~\ref{expt}(e) and (f). With further increase in $R_d$ the oscillation is quenched and the system rests in the stable steady state. Fig.~\ref{expt}(g) and (h) shows the case for $R_d\approx 895$ \ohm. Note the qualitative resemblance between the experimental scenarios and the numerical results of Fig.\ref{vd}.

\section{conclusion}
In summary, we have proposed a scheme to control birhythmic behavior in nonlinear oscillators. Our control scheme incorporates a self-feedback term that is governed by the variable to be controlled and its canonical conjugate. We have considered a prototypical model that shows birhythmic oscillation and has relevance in modeling biochemical processes. Our study has revealed that a proper choice of the control parameter can effectively eliminate birhythmicity for any choice of nonlinear damping parameters and at the same time its variation may give rise to transitions between several interesting dynamical behaviors. Physical implementation of our control scheme is very much feasible, since feedback  through conjugate variables is quite natural in many experimental setups \cite{karna}. We can realize the control scheme if we have access to at least one of the variables of interest; from that we can always generate its time derivative via real time signal processing. We believe that our study may have potential applications in controlling birhythmicity in several mechanical and biochemical processes as well as in other fields.  

\acknowledgments
D.B. acknowledges the financial support from CSIR, New Delhi, India,  T.B. acknowledges the financial support from SERB, Department of Science and Technology (DST), India [Project Grant No.: SB/FTP/PS-005/2013].
\providecommand{\noopsort}[1]{}\providecommand{\singleletter}[1]{#1}%
\end{document}